\documentclass[twocolumn,english,prl,epsfig,rotate,showpacs,aps]{revtex4-2}
\usepackage[T1]{fontenc}
\usepackage[utf8]{inputenc}
\usepackage{color}
\usepackage{epsfig}
\usepackage{verbatim}
\usepackage{amsmath}
\usepackage{amsthm}
\usepackage{amssymb}
\usepackage{graphicx}
\usepackage{esint}
\usepackage{microtype}
\usepackage{dcolumn}
\usepackage{bm}
\usepackage{amsfonts}
\usepackage[table]{xcolor} %
\usepackage{subfigure}

\newcommand{\ReOp}{\operatorname{Re}}

\newcommand{\intk}{\int \frac{d^D k}{(2\pi)^D}}
\makeatletter
\usepackage{babel}
\usepackage{braket}

\makeatother

\usepackage{babel}
\begin{document}
\renewcommand{\figurename}{Fig.}
\title{Quantum Geometry-Driven Nonlinear Spin Currents in Floquet Non-Hermitian Altermagnets} 
\author{Kai Chen$^{\dagger}$}
\author{Jie Zhu$^{*}$}
\affiliation{School of Physics Science and Engineering, Tongji University, Shanghai, China}
\date{\today}
\begin{abstract}
Altermagnets are rapidly emerging as a highly promising platform for spintronics, yet dynamically controlling their spin responses remains a fundamental challenge. In this work, we demonstrate that introducing periodic optical driving and non-Hermiticity provides a powerful route to achieve tunable control over these systems. We derive a general analytical expression for nonlinear spin currents in non-Hermitian phases with a spectral line gap, revealing that the intrinsic response cleanly separates into quantum metric, Berry curvature, and Berry connection dipole contributions. Applying this formalism to a Floquet non-Hermitian $d$-wave altermagnet, we uncover that the nonlinear spin conductivity is overwhelmingly dominated by the bare quantum metric. Furthermore, we show that the optical field's polarization can actively tune---and even strictly reverse---the direction of both longitudinal and transverse spin currents. Our work establishes a quantum geometric framework for the optical manipulation of nonlinear spin transport in advanced magnetic materials.
\end{abstract}

\maketitle

\textit{Introduction.}\textbf{---}Non-Hermiticity opens the door to manipulating new phases of matter and fabricating devices with capabilities fundamentally beyond the paradigm of Hermitian systems \cite{bender2007making,ashida2020non,bergholtz2021exceptional}. A uniquely dominant phenomenon in this regime is the celebrated non-Hermitian skin effect \cite{okuma2020topological,zhang2022universal,ding2022non}, which is intrinsically tied to the point-gap topology of complex energy spectra. Distinct from point gaps, non-Hermitian systems can also feature line-gap topology. Although line-gapped phases can be adiabatically connected to Hermitian systems, their non-Hermitian nature still plays a crucial role in topological classification \cite{kawabata2019symmetry,kawabata2019classification} and the system's response to external fields \cite{chen2026non}. Specifically, non-Hermiticity modifies the action of fundamental symmetries by breaking the equivalence between transposition and complex conjugation \cite{kawabata2019symmetry,chen2022non}. This decouples time-reversal and particle-hole symmetries, thereby broadening the renowned 10-fold Altland-Zirnbauer classification \cite{altland1997nonstandard} into the 38-fold Bernard-LeClair symmetry class\cite{zhou2019periodic,okuma2023non}.

Furthermore, recent research highlights that topological invariants reflect only a portion of the quantum state space geometry The remaining geometric information is captured by the quantum metric (QM), which characterizes the fundamental distance between two adjacent states in Hilbert space \cite{torma2023essay,gao2025quantum,jiang2025revealing,liu2025quantum,chen2025effect}. In topological phases, the QM is fundamentally bounded from below by the Berry curvature and directly dictates the quantum Fisher information \cite{gao2025quantum}. Driven by these geometric insights, QM-dependent electrical and optical responses have recently garnered extensive theoretical and experimental interest in both Hermitian \cite{wang2023quantum,gao2023quantum,yu2025quantum,liu2025giant,fang2024quantum} and non-Hermitian systems \cite{chen2026non,chen2024quantum,sim2023quantum,dong2026non,montag2026quantum}. Specifically, the QM is now understood to fundamentally govern macroscopic phenomena such as the superfluid weight \cite{torma2018quantum,chen2025effect,chen2024ginzburg,iskin2018quantum}, linear and nonlinear optical conductivities \cite{das2023intrinsic,ghosh2024probing,li2026quantum}, and the nonlinear quantum valley Hall effect \cite{das2024nonlinear}. This rapidly expanding landscape demonstrates that the full quantum geometric tensor (QGT)---which unites the QM and the Berry curvature---is not merely a formal mathematical construct, but the central actor driving diverse material responses \cite{gao2025quantum}.

The conventional framework of magnetism recognizes two primary collinear states: ferromagnetism, which globally breaks time-reversal symmetry to produce a net magnetization, and antiferromagnetism, which maintains zero net magnetization through symmetric, anti-aligned sublattices. Recently, rigorous symmetry-based classifications have expanded this framework, formally identifying altermagnetism as a third fundamental phase driven by momentum-dependent spin splitting \cite{vsmejkal2022emerging,mazin2022altermagnetism,song2025altermagnets}. This theoretical breakthrough has been rapidly substantiated by experimental literature. Notably, materials such as the metallic rutile oxide $\mathrm{RuO}_2$ and the semiconducting chalcogenide $\mathrm{MnTe}$ have emerged as  experimental platforms \cite{fedchenko2024observation,guo2024direct,lee2025dichotomous,lee2024broken,jeong2026altermagnetic}. Recent transport measurements and angle-resolved photoemission spectroscopy (ARPES) on these compounds have directly confirmed the hallmark altermagnetic spin splitting, establishing them as ideal foundations for exploring nontrivial spin responses and dynamically driven magnetic phenomena \cite{liu2026symmetry,choi2026exploring}.

Building upon these  material platforms, the theoretical framework of altermagnetism can be powerfully extended into the non-Hermitian regime. Specifically, non-Hermiticity can be deliberately engineered by coupling an altermagnet to an adjacent ferromagnetic layer. The spin-exchange interactions at this interface introduce a complex self-energy into the effective Hamiltonian describing the altermagnet \cite{cayao2023exceptional,correa2026emergent}. In this work, we derive a general analytical framework for evaluating nonlinear spin currents in such line-gapped non-Hermitian systems. We reveal that this intrinsic spin transport is fundamentally governed by the quantum geometric tensor, explicitly separating into distinct contributions from the quantum metric, the Berry curvature, and the Berry connection dipole. 

Furthermore, we note that the application of periodic optical fields to magnetic systems is rapidly expanding. Recent literature has demonstrated that Floquet engineering can dynamically lift Kramers spin degeneracy in conventional collinear antiferromagnets, enabling the generation of steady-state pure spin currents and the realization of light-induced odd-parity magnetic states \cite{huang2025light,zhu2026floquet,li2026floquet}.

To bridge our theoretical framework with these discoveries, we investigate the periodically driven non-Hermitian d-wave altermagnet. While our analytical expression for nonlinear spin currents requires a non-Hermitian phase featuring a spectral line gap, the pristine d-wave altermagnet naturally exhibits gapless nodal structures. Introducing a periodic light field dynamically opens a spectral gap, successfully driving the system into the requisite line-gapped phase. This Floquet engineering not only provides the necessary conditions to directly apply our general formula, but also highlights the broad utility of optical polarization as a , active knob for tuning nontrivial spin responses and manipulating quantum geometry.

\begin{figure}[h]
\includegraphics[width=1.\columnwidth,height=1.\textheight,keepaspectratio]{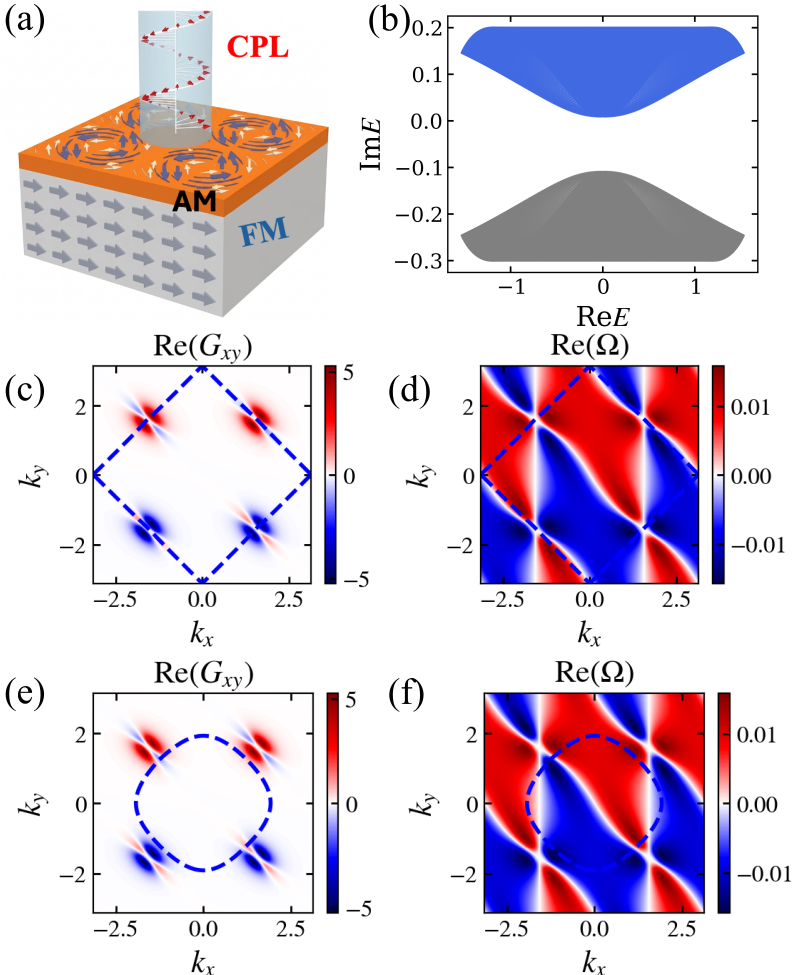}
\caption{(a) Schematic of the light-driven non-Hermitian altermagnet. (b) Complex energy spectrum. (c), (d) Real part of the quantum metric $g_{xy}$ and Berry curvature $\Omega$ at chemical potential $\mu=0$. (e), (f) Same as (c) and (d) for $\mu=-0.5$. Dashed blue lines indicate the Fermi surfaces originating from the blue band in (b). Parameters are set as $t_0=0.5$, $\alpha=0.2$, $\lambda_{\text{R}}=0.01$, $\Gamma=0.05$, $\gamma=0.25$, $\theta=\pi/4$, $\phi=\pi/2$, and $\omega=10$.}
\label{Fig0}
\end{figure}

\textit{Intrinsic nonlinear spin currents (INSC).}---While the nonlinear electric responses of non-Hermitian systems have been extensively studied---particularly uncovering the intrinsic role played by the quantum metric---the nonlinear response of the spin current to an applied electric field, as well as its application to magnetic systems, remains largely unexplored. In this section, we derive an analytical expression for this nonlinear response in a paradigmatic non-Hermitian setup: the line-gapped system, where the energy bands are strictly separated by a line in the complex energy plane. 

Although line-gapped non-Hermitian systems can be adiabatically deformed into Hermitian ones to facilitate topological classification (using established frameworks such as $K$-theory), this mapping inherently discards crucial geometric information. The non-Hermiticity embedded in the complex spectrum and the biorthogonal quantum states imposes a fundamentally distinct geometric structure onto the projective Hilbert space. These modifications are captured by the quantum geometric tensor, which profoundly dictates the system's response to external fields. For a homogeneous non-Hermitian system, the spin current is defined as \cite{culcer2004semiclassical} $\mathbf{J}^s = \int_{\mathbf{k}} f_{\mathbf{k}} \mathbf{s}_n \dot{\mathbf{r}}$, where $\mathbf{s}_n$ is the spin expectation value for the $n$-th state of the unperturbed non-Hermitian system, and the equilibrium Fermi-Dirac distribution takes the explicit form $f_{\mathbf{k}} = \left[ 1 + \exp(\text{Re}\,\xi_{n} / k_B T) \right]^{-1}$, with $\text{Re}\,\xi_{n}$ denoting the real part of the complex eigenenergy for the $n$-th unperturbed band. Here, $T$ is the temperature and $k_B$ is the Boltzmann constant (throughout this work, we set $\hbar=1$, $k_B = 1$, and the elementary charge $e = 1$). The momentum space integration is denoted by $\int_{\mathbf{k}} = \int d^Dk$, where $D$ is the spatial dimension.

Recent studies on non-Hermitian systems have demonstrated that the nonlinear electric current depends sensitively on the product of the wave-packet width and the relaxation time $\tau$. In this work, however, we focus exclusively on the intrinsic nonlinear response of the spin current to an applied electric field. Because this purely geometric contribution is independent of $\tau$, wave-packet width effects can be safely neglected. By applying the Schrieffer-Wolff transformation to the non-Hermitian Hamiltonian perturbed by the electric dipole interaction $-e\mathbf{E}\cdot\mathbf{r}$, we can express the second-order nonlinear spin current as $\mathcal{J}_i^{\alpha(2)} = \sigma_{\alpha}^{i\mu\nu} E^\mu E^\nu$, where the intrinsic nonlinear spin conductivity tensor is given by (for a detailed derivation, see Appendix A):

\begin{equation}
    \sigma^{i\mu\nu}_{\alpha} = \int_{\mathbf{k}} f_{\mathbf{k}} \left( \Gamma^{\text{geom}}_{i\mu\nu} + \Gamma^{\text{magneto}}_{i\mu\nu} + \Gamma^{\text{polar}}_{i\mu\nu} \right),
\label{eqcs}
\end{equation}

The terms $\Gamma^{\text{geom}}_{i\mu\nu}$, $\Gamma^{\text{magneto}}_{i\mu\nu}$, and $\Gamma^{\text{polar}}_{i\mu\nu}$ depend on the band-renormalized quantum metric, the Berry curvature of the unperturbed system, and the Berry connection dipole contributions, respectively. They are explicitly given by:

\begin{equation}
    \Gamma^{\text{geom}}_{i\mu\nu} = e^2 s_{nn}^{\alpha} \mathrm{Re} \left[ 2\partial_i G_{\mu\nu}^{LR} - \frac{\partial_\nu G_{\mu i}^{LR} + \partial_\mu G_{\nu i}^{LR}}{2} \right],
   \label{eqcs1}
\end{equation}

\begin{align}
    \Gamma^{\text{magneto}}_{i\mu\nu} = -\frac{e^2}{2} \sum_{m \neq n} \frac{A_{nm,\mu}^{LR} s^\alpha_{mn} + s^\alpha_{nm} A_{mn,\mu}^{LR}}{\xi_n - \xi_m} \epsilon_{i\nu l} \mathrm{Re}[\Omega_l] \nonumber \\
+ (\mu \leftrightarrow \nu),
\label{eqcs2}
\end{align}

where $n$ labels the band of interest, and $G_{\mu\nu}^{LR}= \sum_{m\ne n}\frac{A_{nm,\mu}^{LR}A_{mn,\nu}^{LR}+A_{nm,\nu}^{LR}A_{mn,\mu}^{LR}}{2(\xi_{n}-\xi_{m})}$ denotes the band-renormalized quantum metric. The term $s^\alpha_{mn} = \langle \psi_{m,\mathbf{k}}^{L} | \hat{s}^\alpha | \psi_{n,\mathbf{k}}^{R} \rangle$ represents the unperturbed biorthogonal spin matrix elements, $A_{mn,\mu}^{LR} = i\langle \psi_{m,\mathbf{k}}^{L} | \partial_\mu \psi_{n,\mathbf{k}}^{R} \rangle$ denotes the non-Hermitian Berry connection.

The first term, $\Gamma^{\text{geom}}_{i\mu\nu}$, is driven by the momentum-space gradients of the non-Hermitian quantum metric $G_{\mu\nu}^{LR}$. Physically, this term captures the real-space shift of the electron wave packet's center of mass during field-induced interband transitions. Because this geometric shift is intrinsically weighted by the unperturbed intraband spin expectation value $s_{nn}^\alpha$, it manifests as a non-Hermitian, spin-polarized analogue of the shift current.

The second term, $\Gamma^{\text{magneto}}_{i\mu\nu}$, represents an anomalous velocity contribution governed by the real part of the Berry curvature $\Omega_l$. Acting as an effective magnetic field in momentum space, the Berry curvature deflects the trajectories of electrons undergoing virtual interband transitions. Intertwined with the biorthogonal spin-mixing matrix elements $s^\alpha_{mn}$, this transverse deflection generates a geometric spin flow akin to an intrinsic, nonlinear spin Hall effect.

Finally, the third term, $\Gamma^{\text{polar}}_{i\mu\nu}$, arises from the gauge-invariant covariant momentum derivative of the Berry connection:

\begin{align}
    \Gamma^{\text{polar}}_{i\mu\nu} &= \frac{e^2}{2} \mathrm{Re}[\partial_i \xi_n] \Bigg\{ i \sum_{m \neq n} \frac{(D_\nu \mathcal{A}_{nm,\mu}^{LR}) s^\alpha_{mn} - s^\alpha_{nm} (D_\nu \mathcal{A}_{mn,\mu}^{LR})}{\xi_n - \xi_m} \nonumber \\
    &\quad + (\mu \leftrightarrow \nu) + \sum_{\substack{m \neq n \\ l \neq n}} \Bigg[ \frac{2 A_{nl,\mu}^{LR} s^\alpha_{lm} A_{mn,\nu}^{LR}}{(\xi_n - \xi_l)(\xi_n - \xi_m)} \nonumber \\
    &\quad - \frac{A_{nm,\mu}^{LR} A_{ml,\nu}^{LR} s^\alpha_{ln} + s^\alpha_{nl} A_{lm,\mu}^{LR} A_{mn,\nu}^{LR}}{(\xi_n - \xi_m)(\xi_m - \xi_l)} \Bigg] \Bigg\},  
 \label{eqcs3}
\end{align}
where $\mathcal{A}_{mn,\mu}^{LR} = \frac{A_{mn,\mu}^{LR}}{\xi_n - \xi_m}$ is the energy-normalized Berry connection, and $D_\nu \mathcal{A}_{mn,\mu}^{LR} \equiv \partial_\nu \mathcal{A}_{mn,\mu}^{LR} - i(A_{nn,\nu}^{LR} - A_{mm,\nu}^{LR})\mathcal{A}_{mn,\mu}^{LR}$ is the gauge-invariant covariant momentum derivative of the energy-normalized Berry connection. This term encapsulates the quantum polarizability, or Berry dipole, of the non-Hermitian bands. The applied electric field dynamically distorts the electron wave packet, modifying its effective group velocity $\mathrm{Re}[\partial_i \xi_n]$. Through multi-band virtual transitions in the non-Hermitian basis, this field-induced polarization yields an additional purely geometric contribution to the macroscopic spin transport.

\begin{figure}[h]
\includegraphics[width=1.\columnwidth,height=1.\textheight,keepaspectratio]{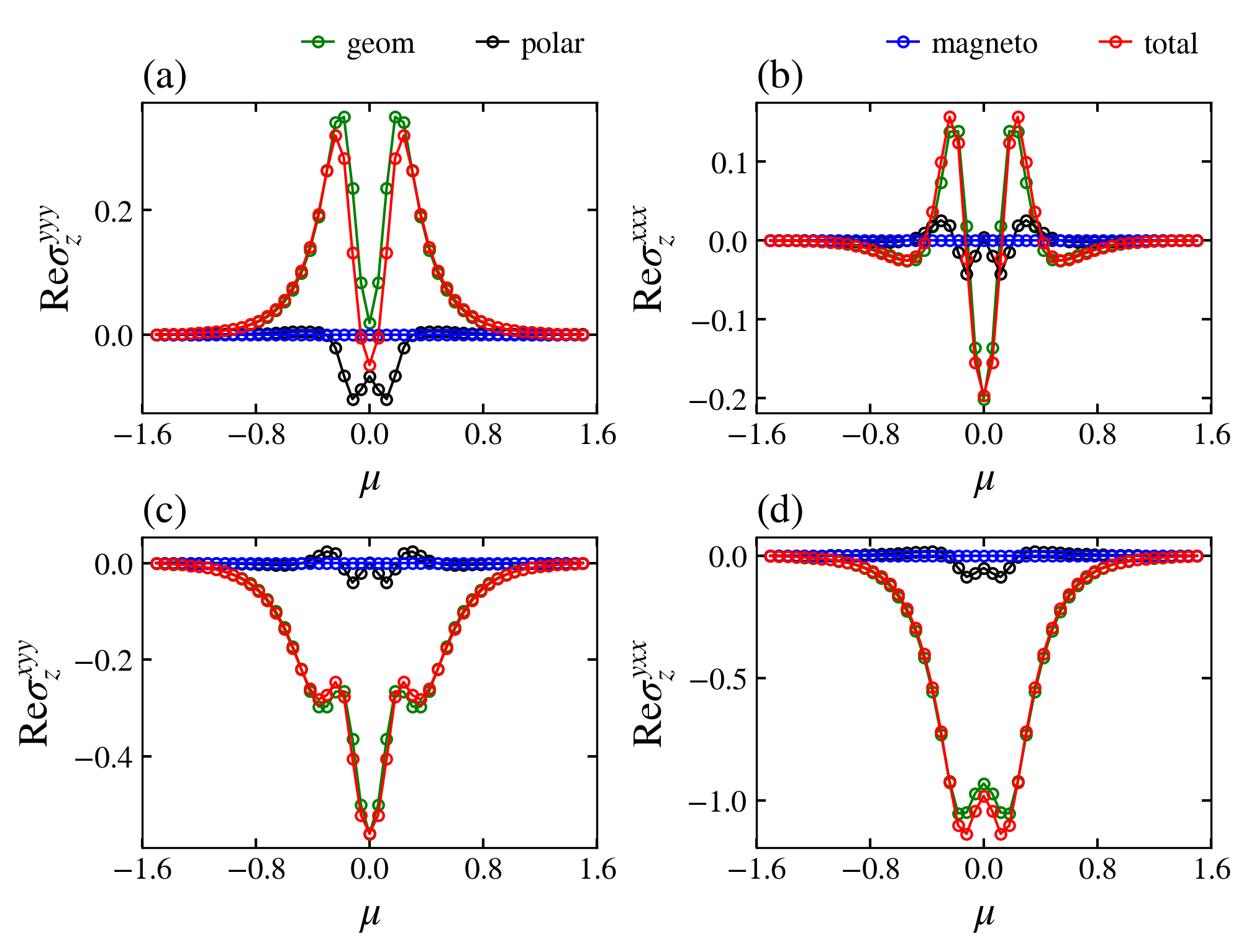}
\caption{Chemical potential dependence of the nonlinear spin conductivity. 
The real part of the nonlinear conductivity tensor components 
(a) $\text{Re}\sigma_{z}^{yyy}$, (b) $\text{Re}\sigma_{z}^{xxx}$, 
(c) $\text{Re}\sigma_{z}^{xyy}$, and (d) $\text{Re}\sigma_{z}^{yxx}$ 
are plotted as a function of the chemical potential $\mu$. 
The total response (red circles) is decomposed into geometric (green), 
polar (black), and magneto (blue) contributions. The numerical results 
are obtained using parameters: $t_0=0.5$, $\alpha=0.2$, $\lambda_R=0.01$, 
$\Gamma=0.05$, $\gamma=0.25$, $\theta=\pi/4$, $\phi=\pi/2$, $\omega=10$, $T=0.01$,
and $A=1$.}
\label{mup}
\end{figure}

\textit{Light-driven non-Hermitian altermagnet.}---To demonstrate the physical implications of our analytical framework, we investigate the nonlinear spin transport in a two-dimensional $d$-wave altermagnet subjected to periodic optical driving. We construct the total Hamiltonian as $H(\mathbf{k}) = H_{0}(\mathbf{k})+H_{\text{AL}}^d(\mathbf{k}) + H_{\text{R}}(\mathbf{k}) + H_{\text{NH}}-\mu\sigma_0$, where non-Hermiticity is introduced via hybridization with an adjacent ferromagnetic layer \cite{cayao2023exceptional,correa2026emergent}. The kinetic term is given by

\begin{equation}
H_{0}(\mathbf{k}) = [-2t_0(\cos k_x + \cos k_y) - \mu] \sigma_0,
\end{equation}
where $t_0$ is the hopping amplitude and $\mu$ is the chemical potential. The $d$-wave altermagnetic core, with its N\'eel vector pinned along the $z$-axis, is described by
\begin{equation}
H_{\text{AL}}^d(\mathbf{k}) = \alpha \left[ 2 \sin k_x \sin k_y \sin 2\theta + (\cos k_x - \cos k_y) \cos 2\theta \right] \sigma_z,
\end{equation}
where $\alpha$ dictates the altermagnetic coupling strength, $\theta$ defines the specific $d$-wave symmetry (e.g., $\theta = 0$ for $d_{x^2-y^2}$ and $\theta = \pi/4$ for $d_{xy}$), and $\sigma_i$ are the Pauli matrices representing the spin degrees of freedom. 

Structural inversion asymmetry and the ferromagnetic proximity effect are captured by the combined Rashba and non-Hermitian Hamiltonian:
\begin{equation}
H_{\text{R}}(\mathbf{k}) + H_{\text{NH}} = \lambda_R (\sin k_y \sigma_x - \sin k_x \sigma_y) - i(\gamma\sigma_x + \Gamma\sigma_0).
\end{equation}
Here, $\lambda_R$ is the Rashba parameter. The non-Hermitian self-energy parameters, $\gamma=(\Gamma_{\uparrow}-\Gamma_{\downarrow})/2$ and $\Gamma=(\Gamma_{\uparrow}+\Gamma_{\downarrow})/2$, arise from the spin-dependent interfacial coupling to an $x$-polarized ferromagnetic lead \cite{cayao2023exceptional,correa2026emergent}.

To dynamically manipulate the system, we introduce an elliptically polarized optical field modeled by the vector potential $\mathbf{A}(t) = A(\sin(\omega t), \cos(\omega t +\phi))$. In the high-frequency regime ($\omega \gg t_0$, where the nearest-neighbor hopping amplitude is set to $t_0 =0.5$), we employ the off-resonant Jacobi-Anger expansion to map the time-periodic system to a static Floquet effective Hamiltonian \cite{goldman2014periodically,eckardt2015high,bukov2015universal}:
\begin{equation}
H_{\text{eff}}(\mathbf{k},\phi) = H(\mathbf{k}) + \frac{1}{\omega} [H_1(\mathbf{k},\phi), H_{-1}(\mathbf{k},\phi)],
\end{equation}
where $H_{\pm 1}(\mathbf{k},\phi)$ are the first-order Fourier harmonics of the driven Hamiltonian. 

Evaluating these commutators yields the effective Hamiltonian in the form
\begin{equation}
H_{\text{eff}}(\mathbf{k},\phi) = \mathbf{h}(\mathbf{k},\phi) \cdot \boldsymbol{\sigma} - (i\Gamma+\mu)\sigma_0,
\label{heff}
\end{equation}
with the detailed analytical expressions for the vector components $h_{x,y,z}(\mathbf{k})$ provided in Appendix B. Crucially, as depicted in Fig.~\ref{Fig0}(b), this driven system exhibits a distinct line gap in the complex energy spectrum. This  spectral separation strictly satisfies the topological prerequisites of our theoretical model, allowing the direct application of the intrinsic nonlinear spin conductivity (INSC) tensor derived previously. Specifically, we restrict the target band index n exclusively to the upper (blue) band. Because the lower (gray) band possesses a significantly more negative imaginary energy, its corresponding states exhibit much shorter lifetimes and decay rapidly as transient modes. Consequently, the long-time steady-state macroscopic transport is effectively governed by the surviving modes of the upper band.

\begin{figure}[h]
\includegraphics[width=1.\columnwidth,height=1.\textheight,keepaspectratio]{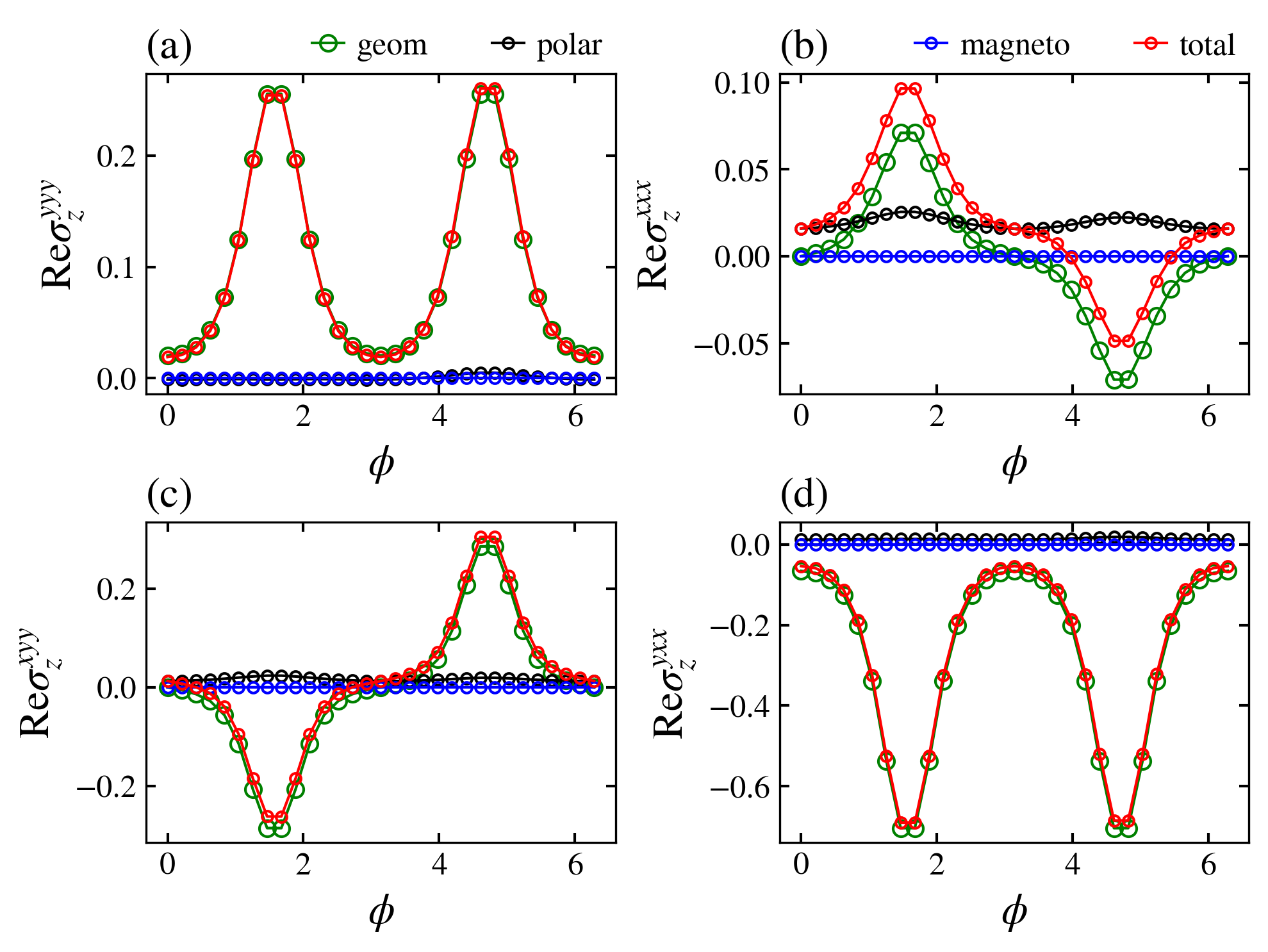}
\caption{Optical field's polarization dependence of the nonlinear spin conductivity. 
The real part of the nonlinear conductivity tensor components 
(a) $\text{Re}\sigma_{z}^{yyy}$, (b) $\text{Re}\sigma_{z}^{xxx}$, 
(c) $\text{Re}\sigma_{z}^{xyy}$, and (d) $\text{Re}\sigma_{z}^{yxx}$ 
are plotted as a function of the chemical potential $\mu$. 
The total response (red circles) is decomposed into geometric (green), 
polar (black), and magneto (blue) contributions. The numerical results 
are obtained using parameters: $t_0=0.5$, $\alpha=0.2$, $\lambda_R=0.01$, 
$\Gamma=0.05$, $\gamma=0.25$, $\theta=\pi/4$, $\mu=-0.3$, $\omega=10$, $T=0.01$,
and $A=1$.}
\label{phi}
\end{figure}

At low temperatures, Eqs.~(\ref{eqcs})--(\ref{eqcs3}) indicate that the spin conductivity depends on the distribution of the QM and the Berry curvature within the Fermi sea. As shown in Figs.~\ref{Fig0}(b)--\ref{Fig0}(f), assuming without loss of generality that the spin current is polarized along the $z$-direction, varying the chemical potential $\mu$ modulates the overlap between the bulk filled states and regions of highly concentrated quantum geometry. As a result, $\mu$ effectively controls the INSCs. This $\mu$-dependence is illustrated in Fig.~\ref{mup}. Both the longitudinal ($\sigma_{z}^{xxx}$ and $\sigma_{z}^{yyy}$) and transverse ($\sigma_{z}^{yxx}$ and $\sigma_{z}^{xyy}$) components are driven by a geometric term related to the quantum metric. Although near $\mu = 0$ these components acquire an additional contribution from the polarization term associated with the Berry curvature dipole, the QM-related term still dominates. Consequently, the bare quantum metric predominantly dictates these nonlinear spin responses.

The effective Hamiltonian in Eq.~(\ref{heff}) depends on the light field's polarization, parameterized by the phase delay $\phi$. Therefore, both the band dispersion and the distribution of the quantum geometric tensor across the Fermi surface and Fermi sea are highly sensitive to the optical polarization. As depicted in Fig.~\ref{phi}, both the longitudinal and transverse INSC components exhibit a strong dependence on $\phi$. The components $\sigma_{z}^{xxx}$ and $\sigma_{z}^{yyy}$ are entirely independent of $\Gamma^{\text{magneto}}$. Most intriguingly, the total response for $\sigma_{z}^{yyy}$, $\sigma_{z}^{xxx}$, $\sigma_{z}^{yxx}$, and $\sigma_{z}^{xyy}$ perfectly tracks the geometric contribution ($\Gamma^{\text{geom}}$), with the longitudinal component $\sigma_{z}^{xxx}$ and transverse component $\sigma_{z}^{xyy}$ exhibiting a strict sign reversal as $\phi$ is varied. Hence, the optical polarization can actively select the direction of both longitudinal and transverse spin currents, paving the way for all-optical manipulation of spin transport in altermagnets. Experimentally, this predicted nonlinear spin transport could be realized in altermagnet/ferromagnet heterostructures (e.g., $\text{RuO}_2$/Permalloy) driven by femtosecond laser pulses, and directly detected via time-resolved terahertz emission spectroscopy.

\textit{Conclusion.---} In summary, we have established a general framework for intrinsic nonlinear spin currents in line-gapped non-Hermitian systems. By applying this to a Floquet $d$-wave altermagnet, we demonstrated that the quantum metric overwhelmingly dominates the macroscopic nonlinear spin conductivity. Furthermore, we showed that the optical field's polarization provides a , active knob to control and reverse these spin currents. Our findings not only deepen the fundamental understanding of non-Hermitian quantum geometry but also pave the way for designing next-generation, all-optically controlled altermagnetic spintronic devices.

\textit{Acknowledgments}\textbf{---} We acknowledge support from the Fundamental Research Funds for the Central Universities. J.Z. acknowledges support from the National Natural Science Foundation of China (Grants No. 92263208), the National Key R\&D Program of China (Grants No. 2022YFA1404400), and the Research Grants Council of Hong Kong SAR (Grant No. AoE/P-502/20).

\noindent $^{\dagger}$: KaiChenPhys@tongji.edu.cn
\noindent $^{*}$: jiezhu@tongji.edu.cn

\bibliographystyle{apsrev4-2}
\bibliography{Lutlib}

\appendix
\onecolumngrid
 
\section*{Appendix for Quantum Geometry-Driven Nonlinear Spin Currents in Floquet Non-Hermitian Altermagnets}

\section{Nonlinear Spin Currents}
To calculate the analytical expressions for the field-perturbed spin expectation values $s_n^{\alpha(1)}$ and $s_n^{\alpha(2)}$, we apply the Schrieffer-Wolff (SW) transformation to the spin operator $\hat{s}^\alpha$. In a non-Hermitian framework, the unperturbed biorthogonal matrix elements of the spin operator are defined using the left and right eigenstates:
\begin{equation}
    s^{\alpha(0)}_{mn} = \langle \psi_{m,\mathbf{k}}^{L} | \hat{s}^\alpha | \psi_{n,\mathbf{k}}^{R} \rangle.
\end{equation}
In the presence of an external electric field $\mathbf{E}$, the quantum states are perturbed. Consequently, any observable evaluated in the $n$-th band is most directly found by transforming the operator itself using the SW generator $S$:
\begin{equation}
    \tilde{s}^\alpha = e^S \hat{s}^\alpha e^{-S} \approx \hat{s}^\alpha + [S, \hat{s}^\alpha] + \frac{1}{2}[S, [S, \hat{s}^\alpha]] + \dots
\end{equation}

To evaluate these commutators, we must first derive the explicit forms of $S$ up to second order in $\mathbf{E}$. The system is governed by the perturbed Hamiltonian $\hat{H} = H_0 - e\mathbf{E}\cdot\hat{\mathbf{r}}$. By expressing the position operator in the momentum-space Bloch basis as $\hat{\mathbf{r}} = i\nabla_\mathbf{k} + \mathbf{A}^{LR}$, we can separate the Hamiltonian into diagonal (intraband) and off-diagonal (interband) components:
\begin{equation}
    \hat{H} = H_0 - e\mathbf{E}\cdot\mathbf{A}_d^{LR} - e\mathbf{E}\cdot\mathbf{A}_{od}^{LR} - ie\mathbf{E}\cdot\nabla_\mathbf{k}.
\end{equation}
The SW transformation block-diagonalizes $\hat{H}$ by eliminating the interband coupling order by order via the expansion of $\tilde{H} = e^S \hat{H} e^{-S}$. 

At first order $\mathcal{O}(E^1)$, eliminating the off-diagonal elements between the target band $n$ and any other band $m \neq n$ imposes the condition $[S^{(1)}, H_0]_{mn} - e\mathbf{E}\cdot\mathbf{A}_{mn}^{LR} = 0$. Evaluating this commutator yields $-S^{(1)}_{mn} (\xi_n^{} - \xi_m^{}) = e\mathbf{E}\cdot\mathbf{A}_{mn}^{LR}$. 

At second order $\mathcal{O}(E^2)$, new off-diagonal terms are generated by the commutation of $S^{(1)}$ with the diagonal field perturbations. Eliminating these cross-terms requires:
\begin{equation}
    [S^{(2)}, H_0]_{mn} + [S^{(1)}, -e\mathbf{E}\cdot\mathbf{A}_d^{LR}]_{mn} + [S^{(1)}, -ie\mathbf{E}\cdot\nabla_\mathbf{k}]_{mn} = 0.
\end{equation}
Evaluating these commutators, the diagonal Berry connection term yields $e^2 E^\mu E^\nu (A_{mm,\nu}^{LR} - A_{nn,\nu}^{LR}) \mathcal{A}_{mn,\mu}^{LR}$, while the momentum derivative acts on the first-order generator to yield $i e^2 E^\mu E^\nu \partial_\nu \mathcal{A}_{mn,\mu}^{LR}$. 

By solving these $\mathcal{O}(E^1)$ and $\mathcal{O}(E^2)$ conditions, the gauge-invariant SW generator $S$ can be separated into its exact first- and second-order components:
\begin{align}
    S_{mn}^{(1)} &= e E^\mu \mathcal{A}_{mn,\mu}^{LR}, \\
    S_{mn}^{(2)} &= -i e^2 E^\mu E^\nu \frac{D_\nu \mathcal{A}_{mn,\mu}^{LR}}{\xi_n^{} - \xi_m^{}},
\end{align}
where $\xi_m^{}$ denotes the complex eigenenergy of the $m$-th unperturbed band (used interchangeably with $\xi_m$ to simplify notation where no confusion arises). We have defined the energy-normalized Berry connection $\mathcal{A}_{mn,\mu}^{LR} = \frac{A_{mn,\mu}^{LR}}{\xi_n^{} - \xi_m^{}}$ and its gauge-invariant covariant momentum derivative $D_\nu \mathcal{A}_{mn,\mu}^{LR} = \partial_\nu \mathcal{A}_{mn,\mu}^{LR} - i(A_{mm,\nu}^{LR} - A_{nn,\nu}^{LR})\mathcal{A}_{mn,\mu}^{LR}$. Furthermore, we adopt a gauge choice such that the intraband generator vanishes ($S_{nn} = 0$), and Einstein summation over repeated spatial indices $\mu, \nu$ is implied throughout.

\subsection{The First-Order Spin Correction $s_n^{\alpha(1)}$}

The linear response of the spin expectation value is given by the commutator with the first-order SW generator evaluated for the target band $n$: 
\begin{equation}
    s_n^{\alpha(1)} = [S^{(1)}, \hat{s}^\alpha]_{nn} = \sum_{m \neq n} \left( S_{nm}^{(1)} s_{mn}^\alpha - s_{nm}^\alpha S_{mn}^{(1)} \right)
\end{equation}
Expanding this using the explicit form of $S_{mn}^{(1)}$, we obtain:
\begin{equation}
    s_n^{\alpha(1)} = -e E^\mu \sum_{m \neq n} \frac{A_{nm,\mu}^{LR} s_{mn}^\alpha + s_{nm}^\alpha A_{mn,\mu}^{LR}}{\xi_n^{} - \xi_m^{}}
\end{equation}
This represents the linear spin polarization induced by the electric field mixing the unperturbed bands, acting as the non-Hermitian equivalent of the Edelstein (inverse spin galvanic) effect.

\subsection{The Second-Order Spin Correction $s_n^{\alpha(2)}$}

The second-order correction contains two parts: the commutator with the second-order generator, and the double commutator with the first-order generator:
\begin{equation}
    s_n^{\alpha(2)} = [S^{(2)}, \hat{s}^\alpha]_{nn} + \frac{1}{2}[S^{(1)}, [S^{(1)}, \hat{s}^\alpha]]_{nn}
\end{equation}
By substituting the gauge-invariant SW generators and utilizing the symmetry of $E^\mu E^\nu$ to cancel the $l=n$ terms in the double commutator, we obtain the full analytical expression:
\begin{align}
    s_n^{\alpha(2)} &= e^2 E^\mu E^\nu \Bigg\{ -i \sum_{m \neq n} \frac{(D_\nu \mathcal{A}_{nm,\mu}^{LR}) s^\alpha_{mn} - s^\alpha_{nm} (D_\nu \mathcal{A}_{mn,\mu}^{LR})}{\xi_n^{} - \xi_m^{}} \notag \\
    &\quad + \frac{1}{2} \sum_{\substack{m \neq n \\ l \neq n}} \left[ \frac{2 A_{nl,\mu}^{LR} s^\alpha_{lm} A_{mn,\nu}^{LR}}{(\xi_n^{} - \xi_l^{})(\xi_n^{} - \xi_m^{})} - \frac{A_{nm,\mu}^{LR} A_{ml,\nu}^{LR} s^\alpha_{ln} + s^\alpha_{nl} A_{lm,\mu}^{LR} A_{mn,\nu}^{LR}}{(\xi_n^{} - \xi_m^{})(\xi_m^{} - \xi_l^{})} \right] \Bigg\}
\end{align}

The first sum (over $m$) represents the geometric dispersion of the transition dipole. By utilizing the covariant derivative $D_\nu$, it correctly captures how the first-order mixing between bands changes across momentum space in a strictly gauge-invariant manner. The second sum (the double sum over $m, l$) is the purely off-diagonal virtual transition term. The electric field couples the target band $n$ through two intermediate bands before evaluating the spin. This structure mathematically mirrors the band-renormalized non-Hermitian quantum metric $G_{\mu\nu}^{LR}$, but with the spin operator inserted into the transition path.

\subsection{Integration into the Semiclassical Spin Current}

Having derived the relevant geometric contributions, the total intrinsic second-order DC spin current $\mathcal{J}^{\alpha(2)}_{\text{int}}$ can be expressed exactly as the sum of three cross-terms arising from the perturbative expansion of the spin and velocity within the semiclassical Boltzmann framework:
\begin{equation}
    \mathcal{J}^{\alpha(2)}_{\text{int}} = \int_\mathbf{k} f_{\mathbf{k}} \left( s_n^{\alpha} \dot{\mathbf{r}}_n^{(2)} + s_n^{\alpha(1)} \dot{\mathbf{r}}_n^{(1)} + s_n^{\alpha(2)} \dot{\mathbf{r}}_n^{(0)} \right).
\end{equation}
This formulation captures not only the anomalous geometric transport of the equilibrium spin, but also how the quantum geometric tensor dictates the transport of the field-induced spin polarization. As demonstrated in Ref.~\cite{chen2026non}, the perturbative corrections to the wave-packet velocity are given by:
\begin{equation} 
\left\{ 
\begin{aligned}
     \dot{\mathbf{r}}_n^{(0)} &= \mathrm{Re} \left[ \nabla_{\mathbf{k}}\xi^{}_{\mathbf{k}} \right], \\
     \dot{\mathbf{r}}_n^{(1)} &= \mathrm{Re} \left[ e\mathbf{E} \times \boldsymbol{\Omega}^{}_\mathbf{k} \right], \\
     \dot{\mathbf{r}}_n^{(2)} &= \mathrm{Re} \left[ \nabla_{\mathbf{k}}\xi^{(2)}_{\mathbf{k}} + e\mathbf{E} \times \boldsymbol{\Omega}^{(1)}_\mathbf{k} \right],  
\end{aligned} 
\right. 
\end{equation}
where $\xi^{(2)}_{\mathbf{k}} = e^2\sum_{\mu,\nu} G^{LR}_{n,\mu\nu}E^{\mu}E^{\nu}$ is the second-order wave-packet energy shift, and $\Omega_{n,\alpha}^{(1)} = -2e\epsilon_{\alpha\beta\nu}\partial_{\beta}G_{n,\mu\nu}^{LR}E^{\mu}$ is the first-order field-induced Berry curvature.

The intrinsic second-order spin conductivity tensor $\sigma_{i\mu\nu}^{\alpha, \text{int}}$ dictates the second-order spin current response $\mathcal{J}_i^{\alpha(2)} = \sigma_{\alpha}^{i\mu\nu} E^\mu E^\nu$. This tensor is perfectly symmetrized with respect to the spatial indices of the electric field ($\mu, \nu$). 

The total intrinsic nonlinear spin conductivity is the sum of three distinct integrands over the Brillouin zone, weighted by the equilibrium Fermi-Dirac distribution $f_{\mathbf{k}}$:
\begin{equation}
    \sigma^{i\mu\nu}_{\alpha} = \intk f_{\mathbf{k}} \left( \Gamma^{\text{geom}}_{i\mu\nu} + \Gamma^{\text{magneto}}_{i\mu\nu} + \Gamma^{\text{polar}}_{i\mu\nu} \right)
\end{equation}

Below are the exact, unabridged expressions for each component derived via the Schrieffer-Wolff transformation.

\subsubsection{The Geometric Spin Transport Term ($\Gamma^{\text{geom}}_{i\mu\nu}$)}

This term captures the unperturbed equilibrium spin $s_{nn}^{\alpha}$ being transported by the second-order anomalous velocity. It is governed directly by the band-renormalized non-Hermitian quantum metric $G_{\mu\nu}^{LR}$:
\begin{equation}
    \Gamma^{\text{geom}}_{i\mu\nu} = e^2 s_{nn}^{\alpha} \ReOp \left[ 2\partial_i G_{\mu\nu}^{LR} - \frac{\partial_\nu G_{\mu i}^{LR} + \partial_\mu G_{\nu i}^{LR}}{2} \right]
\end{equation}

\subsubsection{The Magnetoelectric Spin Hall Term ($\Gamma^{\text{magneto}}_{i\mu\nu}$)}

This term captures the first-order, field-induced spin polarization (the inverse spin galvanic effect) being pushed transversely by the first-order anomalous Hall velocity. It depends on the unperturbed Berry curvature $\Omega_l^{}$:
\begin{equation}
    \Gamma^{\text{magneto}}_{i\mu\nu} = -\frac{e^2}{2} \sum_{m \neq n} \left( \frac{A_{nm,\mu}^{LR} s^\alpha_{mn} + s^\alpha_{nm} A_{mn,\mu}^{LR}}{\xi_n^{} - \xi_m^{}} \epsilon_{i\nu l} \ReOp[\Omega_l^{}] + (\mu \leftrightarrow \nu) \right)
\end{equation}

\subsubsection{ The Second-Order Polarization Term ($\Gamma^{\text{polar}}_{i\mu\nu}$)}

This term represents the second-order spin polarization being carried longitudinally by the unperturbed group velocity. It requires the full gauge-invariant second-order Schrieffer-Wolff generator, accounting for both the covariant dispersion of the interband dipole and the off-diagonal double-commutator virtual transitions:
\begin{align}
    \Gamma^{\text{polar}}_{i\mu\nu} &= \frac{e^2}{2} \ReOp[\partial_i \xi_n^{}] \Bigg\{- i \sum_{m \neq n} \frac{(D_\nu \mathcal{A}_{nm,\mu}^{LR}) s^\alpha_{mn} - s^\alpha_{nm} (D_\nu \mathcal{A}_{mn,\mu}^{LR})}{\xi_n^{} - \xi_m^{}} + (\mu \leftrightarrow \nu) \nonumber \\
    &\quad + \sum_{\substack{m \neq n \\ l \neq n}} \left[ \frac{2 A_{nl,\mu}^{LR} s^\alpha_{lm} A_{mn,\nu}^{LR}}{(\xi_n^{} - \xi_l^{})(\xi_n^{} - \xi_m^{})} - \frac{A_{nm,\mu}^{LR} A_{ml,\nu}^{LR} s^\alpha_{ln} + s^\alpha_{nl} A_{lm,\mu}^{LR} A_{mn,\nu}^{LR}}{(\xi_n^{} - \xi_m^{})(\xi_m^{} - \xi_l^{})} \right] \Bigg\}
\end{align}

\section{Floquet Effective Hamiltonian}
\label{app:floquet}

\subsection{The Model Hamiltonian}

We consider a two-dimensional $d$-wave altermagnet (AM) on a square lattice. The system lacks inversion symmetry, which gives rise to Rashba spin-orbit coupling. The total Hamiltonian is expressed as $H(\mathbf{k}) = H_0(\mathbf{k}) + H_{\text{AL}}^d(\mathbf{k}) + H_{\text{R}}(\mathbf{k})$.

The kinetic term is given by
\begin{equation}
H_{0}(\mathbf{k}) = [-2t_0(\cos k_x + \cos k_y) - \mu] \sigma_0,
\end{equation}
where $t_0$ is the hopping amplitude and $\mu$ is the chemical potential. The $d$-wave altermagnetic contribution, with the N\'{e}el vector oriented along the $z$ axis, is 
\begin{equation}
H_{\text{AL}}^d(\mathbf{k}) = \alpha \left[ 2 \sin k_x \sin k_y \sin 2\theta + (\cos k_x - \cos k_y) \cos 2\theta \right] \sigma_z,
\end{equation}
where $\alpha$ denotes the altermagnetic coupling strength. The angle $\theta$ parameterizes the symmetry of the order parameter; for instance, $\theta = 0$ and $\theta = \pi/4$ correspond to $d_{x^2-y^2}$ and $d_{xy}$ symmetries, respectively.

The Rashba SOC Hamiltonian is:
\begin{equation}
H_R(\mathbf{k}) = \lambda_R (\sin k_y \sigma_x - \sin k_x \sigma_y)
\end{equation}

\subsection{Peierls Substitution and the Light Field}
To introduce a periodic driving light field, we use an elliptically polarized vector potential:
\begin{equation}
\mathbf{A}(t) = A (\sin \omega t, \cos(\omega t + \phi))
\end{equation}
where $A = E_0/\omega$ is the dimensionless light amplitude (assuming $e = \hbar = a = 1$). Under the Peierls substitution, the crystal momenta become time-dependent:
\begin{equation}
k_x(t) = k_x + A \sin(\omega t), \quad k_y(t) = k_y + A \cos(\omega t + \phi)
\end{equation}

To handle the diagonal hopping terms ($\sin k_x \sin k_y$) in the Floquet expansion, we rewrite them using sum/difference formulas:
\begin{equation}
k_x(t) \pm k_y(t) = k_x \pm k_y + A_\pm \sin(\omega t + \gamma_\pm)
\end{equation}
where the modified amplitudes $A_\pm$ and phase shifts $\gamma_\pm$ are defined as:
\begin{equation}
A_{\pm} = A \sqrt{2 \mp 2\sin \phi}, \quad \cos \gamma_{\pm} = \frac{1 \mp \sin \phi}{\sqrt{2 \mp 2\sin \phi}}, \quad \sin \gamma_{\pm} = \frac{\pm \cos \phi}{\sqrt{2 \mp 2\sin \phi}}
\end{equation}

\subsection{High-Frequency Floquet Effective Hamiltonian ($H_{eff}$)}
Using the Jacobi-Anger expansion, $e^{i z \sin(\omega t + \delta)} = \sum_l J_l(z) e^{il(\omega t + \delta)}$, we map the periodically driven system to an effective static Hamiltonian in the high-frequency limit ($\omega \gg t_0$):
\begin{equation}
H_{eff}(\mathbf{k}) = H(\mathbf{k}) + \frac{1}{\omega} [H_1(\mathbf{k}), H_{-1}(\mathbf{k})]
\end{equation}

By extracting the zeroth and first-order ($\pm 1$) Fourier harmonics and evaluating the commutators, we define the effective Hamiltonian as a sum of Pauli matrices:
\begin{equation}
H_{eff}(\mathbf{k}) = h_x(\mathbf{k}) \sigma_x + h_y(\mathbf{k}) \sigma_y + h_z(\mathbf{k}) \sigma_z-(i\Gamma+\mu)\sigma_0
\end{equation}

The vector components of the effective Hamiltonian are analytically found to be:
\begin{align}
h_x(\mathbf{k}) &= \lambda_R J_0(A) \sin k_y - \frac{4\lambda_R}{\omega} J_1(A) \cos k_x \text{Re}[A_1(\mathbf{k})] -i\gamma\\
h_y(\mathbf{k}) &= -\lambda_R J_0(A) \sin k_x - \frac{4\lambda_R}{\omega} J_1(A) \cos k_y \text{Im}[A_1(\mathbf{k}) e^{-i\phi}] \\
h_z(\mathbf{k}) &= H_{0, AL}^z(\mathbf{k}) + \frac{4\lambda_R^2}{\omega} J_1^2(A) \cos\phi \cos k_x \cos k_y
\end{align}
where $H_{0, AL}^z(\mathbf{k})$ is the time-averaged $d$-wave term:
\begin{equation}
H_{0, AL}^z(\mathbf{k}) = \alpha \left[ \big( J_0(A_-) \cos(k_x-k_y) - J_0(A_+) \cos(k_x+k_y) \big)\sin 2\theta + J_0(A)(\cos k_x - \cos k_y)\cos 2\theta \right]
\end{equation}
and the complex scalar factor $A_1(\mathbf{k})$ generating the cross-commutator fields is defined by its real and imaginary parts:
\begin{align}
\text{Re}[A_1(\mathbf{k})] &= \alpha \Big[ \big( J_1(A_+)\sin\gamma_+ \sin(k_x+k_y) - J_1(A_-)\sin\gamma_- \sin(k_x-k_y) \big)\sin 2\theta \nonumber \\
&\quad\quad + J_1(A)\cos\phi \sin k_y \cos 2\theta \Big] \\
\text{Im}[A_1(\mathbf{k}) e^{-i\phi}] &= \alpha \Big[ \big( J_1(A_-)\cos(\gamma_--\phi)\sin(k_x-k_y) - J_1(A_+)\cos(\gamma_+-\phi)\sin(k_x+k_y) \big)\sin 2\theta \nonumber \\
&\quad\quad + J_1(A)\cos\phi \sin k_x \cos 2\theta \Big]
\end{align}

\end{document}